\documentclass{aa}
\usepackage{natbib}
\usepackage{epic,eepic}

\begin{document}
 
%
%
\def\valid{}    

\newcommand{\bea}{\begin{eqnarray}}
\newcommand{\eea}{\end{eqnarray}}

\font\caps=cmcsc10                  
\font\dunh=cmdunh10  at 12.0 true pt 
\font\dunhs=cmdunh10 
\font\vbold=cmbx10 scaled \magstep1 
\font\sevenbf=cmbx7
\font\sevenit=cmti7
\font\Kapi=cmr17

\def\MEV{DOME}
\def\RTE{equation of radiative transfer}
\def\etal{{et al}}
\def\HW{H\&W}
\def\OK{O\&K}
\def\ok{O\&K}
\def\RH{R\&H}

\def\ibmrs{\hbox{\tt RS/6000}}
\def\hp{\hbox{\tt HP~9000}}
\def\dec{\hbox{\tt DEC~5000}}
\def\axp{\hbox{\tt AXP}}
\def\ibmmf{\hbox{\tt IBM~3090}}
\def\ibmpc{\hbox{\tt 486DX}}
\def\cray{\hbox{\tt Cray 2}}
\def\ymp{\hbox{\tt YMP}}
\def\nec{\hbox{\tt NEC}}

\def\g{\gamma}
\def\b{\beta}
\def\m{\mu}
\def\e{\epsilon}
\def\n{\nu}
\def\l{\lambda}
\def\L{\Lambda}
\def\t{\tau}
\def\pder#1#2{{\partial #1 \over \partial #2}}
\def\div#1#2{{#1\over #2}}
\def\rout{\ifmmode{r_{\rm out}}\else\hbox{$r_{\rm out}$}\fi}
\def\tmax{\ifmmode{\tau_{\rm max}}\else\hbox{$\tau_{\rm max}$}\fi}
\def\tstd{\ifmmode{\tau_{\rm std}}\else\hbox{$\tau_{\rm std}$}\fi}
\def\vmax{\ifmmode{v_{\rm max}}\else\hbox{$v_{\rm max}$}\fi}
\def\muE{\ifmmode{\mu_{\rm E}}\else\hbox{$\mu_{\rm E}$}\fi} 
\def\pE{\ifmmode{p_{\rm E}}\else\hbox{$p_{\rm E}$}\fi} 
\def\bmax{\ifmmode{\b_{\rm max}}\else\hbox{$\b_{\rm max}$}\fi}
\def\kms{\hbox{$\,$km$\,$s$^{-1}$}}
\def\ergs{\hbox{$\,$erg$\,$s$^{-1}$}}
\def\kpc{\hbox{$\,$kpc} }
\def\ang{\hbox{\AA}}
\def\Msun{\hbox{$\,$M$_\odot$} }
\def\Lsun{\hbox{$\,$L$_\odot$} }
\def\Teff{\hbox{$\,T_{\rm eff}$} }
\def\alog#1{\times 10^{#1}}
\def\rin{\hbox{$r_{\rm in}$} }
\def\rout{\hbox{$r_{\rm out}$} }

\def\lstar{\ifmmode{\Lambda^*}\else\hbox{$\Lambda^*$}\fi} 
\def\Lstar{\ifmmode{\Lambda^*}\else\hbox{$\Lambda^*$}\fi} 
\def\Rop{\ifmmode{[R_{ij}]}\else\hbox{$[R_{ij}]$}\fi}
\def\Rij{\Rop}
\def\Rji{\ifmmode{[R_{ji}]}\else\hbox{$[R_{ji}]$}\fi}
\def\Rstar{\ifmmode{[R_{ij}^*]}\else\hbox{$[R_{ij}^*]$}\fi}
\def\Rijstar{\Rstar}
\def\Rjistar{\ifmmode{[R_{ji}^*]}\else\hbox{$[R_{ji}^*]$}\fi}
\def\DRji{\ifmmode{[\Delta R_{ji}]}\else\hbox{$[\Delta R_{ji}]$}\fi}
\def\DRij{\ifmmode{[\Delta R_{ij}]}\else\hbox{$[\Delta R_{ij}]$}\fi}

\def\Jb{{ J}}
\def\Jnew{{ J_{\rm new}}}
\def\Jold{{ J_{\rm old}}}
\def\Jfs{{ J_{\rm fs}}}
\def\Snew{{S_{\rm new}}}
\def\Sold{{S_{\rm old}}}
\def\Amat{\mat{A}}             

\def\ns{\ifmmode{N_{\rm s}}          
        \else\hbox{$N_{\rm s}$}\fi}
\def\ion#1{\hbox{ #1}}         

\def\peq{\mathbin{\hbox{$+$}\hbox{$=$}}}

\def\mat#1{{\bf #1}}     
\def\vek#1{{#1}}         

\newcount\eqcount
\eqcount=0
\def
  \nummer{
    \global\advance\eqcount by 1
    (\the\eqcount)
  }

\def
  \numadv{
    \global\advance\eqcount by 1
  }

\def
   \numout#1{
     (\the\eqcount #1)
  }

\def\ivek#1#2{\ifmmode{\vek{I}^{#1}_{#2}}
        \else\hbox{$\vek{I}^{#1}_{#2}$}\fi}

\def\ip#1{\ivek{+}{#1}}      
\def\im#1{\ivek{-}{#1}}      

\def\tmat#1#2{\ifmmode{\mat{t}^{#1}_{#2}}
        \else\hbox{$\mat{t}^{#1}_{#2}$}\fi}
\def\rmat#1#2{\ifmmode{\mat{r}^{#1}_{#2}}
        \else\hbox{$\mat{r}^{#1}_{#2}$}\fi}
\def\bvek#1#2{\ifmmode{\beta^{#1}_{#2}}
        \else\hbox{$\beta^{#1}_{#2}$}\fi}

\def\tpi#1{\tmat{+}{#1}}
\def\tmi#1{\tmat{-}{#1}}
\def\rmi#1{\rmat{-}{#1}}
\def\rpi#1{\rmat{+}{#1}}
\def\bpi#1{\bvek{+}{#1}}
\def\bmi#1{\bvek{-}{#1}}

\def\tp{\tmat{+}{}}          
\def\tm{\tmat{-}{}}          
\def\rmm{\rmat{-}{}}         
\def\rp{\rmat{+}{}}          
\def\bp{\bvek{+}{}}          
\def\bm{\bvek{-}{}}          
\def\tpm{\tmat{\pm}{}}       
\def\rpm{\rmat{\pm}{}}       
\def\bpm{\bvek{\pm}{}}       

\def\lp{\ifmmode{\lambda^+_\tau}           
        \else\hbox{$\lambda^+_\tau$}\fi}
\def\lm{\ifmmode\lambda^-_\tau             
        \else\hbox{$\lambda^-_\tau$}\fi}

%
%
%
%



\def\aasref@jnl#1{{\rm #1}}

\def\aj{\aasref@jnl{AJ}}                   
\def\araa{\aasref@jnl{ARA\&A}}             
\def\apj{\aasref@jnl{ApJ}}                 
\def\apjl{\aasref@jnl{ApJ}}                
\def\apjs{\aasref@jnl{ApJS}}               
\def\ao{\aasref@jnl{Appl.~Opt.}}           
\def\apss{\aasref@jnl{Ap\&SS}}             
\def\aap{\aasref@jnl{A\&A}}                
\def\aapr{\aasref@jnl{A\&A~Rev.}}          
\def\aaps{\aasref@jnl{A\&AS}}              
\def\azh{\aasref@jnl{AZh}}                 
\def\baas{\aasref@jnl{BAAS}}               
\def\jrasc{\aasref@jnl{JRASC}}             
\def\memras{\aasref@jnl{MmRAS}}            
\def\mnras{\aasref@jnl{MNRAS}}             
\def\pra{\aasref@jnl{Phys.~Rev.~A}}        
\def\prb{\aasref@jnl{Phys.~Rev.~B}}        
\def\prc{\aasref@jnl{Phys.~Rev.~C}}        
\def\prd{\aasref@jnl{Phys.~Rev.~D}}        
\def\pre{\aasref@jnl{Phys.~Rev.~E}}        
\def\prl{\aasref@jnl{Phys.~Rev.~Lett.}}    
\def\pasp{\aasref@jnl{PASP}}               
\def\pasj{\aasref@jnl{PASJ}}               
\def\qjras{\aasref@jnl{QJRAS}}             
\def\skytel{\aasref@jnl{S\&T}}             
\def\solphys{\aasref@jnl{Sol.~Phys.}}      
\def\sovast{\aasref@jnl{Soviet~Ast.}}      
\def\ssr{\aasref@jnl{Space~Sci.~Rev.}}     
\def\zap{\aasref@jnl{ZAp}}                 
\def\nat{\aasref@jnl{Nature}}              
\def\iaucirc{\aasref@jnl{IAU~Circ.}}       
\def\aplett{\aasref@jnl{Astrophys.~Lett.}} 
\def\apspr{\aasref@jnl{Astrophys.~Space~Phys.~Res.}}
\def\bain{\aasref@jnl{Bull.~Astron.~Inst.~Netherlands}} 
\def\fcp{\aasref@jnl{Fund.~Cosmic~Phys.}}  
\def\gca{\aasref@jnl{Geochim.~Cosmochim.~Acta}}   
\def\grl{\aasref@jnl{Geophys.~Res.~Lett.}} 
\def\jcp{\aasref@jnl{J.~Chem.~Phys.}}      
\def\jgr{\aasref@jnl{J.~Geophys.~Res.}}    
\def\jqsrt{\aasref@jnl{J.~Quant.~Spec.~Radiat.~Transf.}}
\def\memsai{\aasref@jnl{Mem.~Soc.~Astron.~Italiana}}
\def\nphysa{\aasref@jnl{Nucl.~Phys.~A}}   
\def\physrep{\aasref@jnl{Phys.~Rep.}}   
\def\physscr{\aasref@jnl{Phys.~Scr}}   
\def\planss{\aasref@jnl{Planet.~Space~Sci.}}   
\def\procspie{\aasref@jnl{Proc.~SPIE}}   

\let\astap=\aap
\let\apjlett=\apjl
\let\apjsupp=\apjs
\let\applopt=\ao

\renewcommand{\textfraction}{0.5}

\title{A 3D radiative transfer framework: I. non-local
operator splitting and continuum scattering problems}

\author{Peter H. Hauschildt\inst{1} and E.~Baron\inst{1,2,3}}

\institute{
Hamburger Sternwarte, Gojenbergsweg 112, 21029 Hamburg, Germany;
yeti@hs.uni-hamburg.de 
\and
Homer L.~Dodge Dept.~of Physics and Astronomy, University of
Oklahoma, 440 W.  Brooks, Rm 100, Norman, OK 73019-2061 USA;
baron@nhn.ou.edu
\and
CRD, Lawrence Berkeley National Laboratory, MS
50F-1650, 1 Cyclotron Rd, Berkeley, CA 94720-8139 USA
}

\date{received 18/July/2005, accepted 03/Jan/2006}
\maketitle

\abstract{We describe a highly flexible framework to solve 3D radiation
transfer problems in scattering dominated environments based on a long
characteristics piece-wise parabolic formal solution and an operator splitting
method. We find that the linear systems are efficiently solved with iterative
solvers such as Gauss-Seidel and Jordan techniques. We use a sphere-in-a-box
test model to compare the 3D results to 1D solutions in order to assess the
accuracy of the method. We have implemented the method for static media,
however, it can be used to solve problems in the Eulerian-frame for media with
low velocity fields.
}

\section{Introduction}

With the increase in computer power in the last few years, 3D
hydrodynamical calculations are becoming increasingly common in
astrophysics. Most hydrodynamical calculations treat radiation in a
simplified manner, since a full solution of the 3-D non-LTE radiative
transfer problem is numerically much more expensive than the hydrodynamical
calculation itself, which already stretch the limits of modern
parallel computers. In many instances, such as the thermonuclear
explosion of a white dwarf (thought to be the progenitor of Type Ia
supernova), the goal of the hydrodynamical simulations is to
understand the mode of combustion and to handle the effects of
turbulence in as realistic manner as possible and radiative transfer
effects are ignored. However, in general since the ultimate validation
or falsification of the results of sophisticated hydrodynamical
modeling will be via comparison with the observed radiation from the
astrophysical object being studied, and the radiation strongly affects
the physical state of the matter in the atmosphere of the object
(where the observed radiation originates) the effect of detailed
radiative transfer effects cannot be ignored.

In many multi-dimensional hydrodynamics codes \cite[for
example ZEUS3D,][]{ZEUS3D}, radiative transfer is treated in a simplified matter
in order to determine the amount of energy transfered between the
matter and radiation (the cooling function) although \citet{DKC96}
presented a full time-dependent 2-D NLTE radiative transfer code,
based on a variable Eddington factor method and equivalent two level
atom formulation.

Recently \citet{RPDM05} presented a method of including 3-D radiative
transfer into modern 3-D hydrodynamical codes (such as the ASCI FLASH
code) but they only treated the solution of the radiative
transfer equation in the absence of scattering, that is they their
method only treats the formal solution of the radiative transfer
equation and not the full self consistent scattering problem where the
right hand side of the radiative transfer problem involves the
radiation field itself. However, in astrophysical systems, the effect
of scattering cannot be ignored and in fact it is due to scattering
that the radiation field decouples from the local emission and absorption
and the effects of the existence of the boundary are communicated
globally over the atmosphere \citep{mihalas78sa}. It is just this
strong non-locality of the radiation field that makes the solution of
the generalized radiative transfer problem so computationally
demanding.

\citet{steiner91} presented a 2-D multi-grid method based on the
short-characteristics method \citep{ok87,OAB} and showed that it worked in the
case of a purely absorptive atmosphere, i.e., that the formal solution was
tractable. \citet{vath94} presented a 3-D short characteristics method and
showed that it had adequate scaling on a SIMD parallel architecture.
{In a series of papers 3-D, short characteristics methods for disk
systems were presented by
\cite{adam90,hummel94a,hummel94b,hummeldachs92,papkalla95}. Our method is
similar in spirit to these works, but we present a more detailed description of
the construction of the approximate lambda operator (ALO), and the method of
solution of the scattering problem.}
 
\citet{BTB97} presented a multi-level, multi-grid, multi-dimensional
radiative transfer scheme, using a lower triangular ALO and solving the
scattering problem via a Gauss-Seidel method. 

\citet*{vannoort02} presented a method of solving the full NLTE
radiative transfer problem using the short characteristics method in
2-D for Cartesian, spherical, and cylindrical geometry. They also used
the technique of accelerated lambda iteration (ALI)
\citep{OAB,ok87}, however they restricted themselves to the case of a
diagonal ALO. In addition they considered the 
case including velocity fields, but their method is feasible only in the 
case of a small velocity gradient across the atmosphere.

We describe below a simple framework to solve three dimensional (3D)
radiation transport (RT) problems with a non-local operator splitting
method. In subsequent papers we will extend this framework to solve 3D
RT problems in relativistically moving configurations. Our method is
similar to those described above, although we consider both short
characteristics and long characteristics methods for the formal
solution of the radiative transfer equation.  We show that
long-characteristics produce a significantly better numerical solution
in our test cases for a strongly
scattering dominated atmosphere. Short-characteristics are known to be
diffusive and it is apparent in our numerical results. We also
implement a partial parallelization of the method, although we defer a
full discussion of the parallelization to future work.

\section{Method}

In the following discussion we use  notation of \citet{s3pap}.
First, we will describe the process for the formal solution, then we will
describe how we construct the non-local approximate $\L$ operator, $\lstar$,
and methods to solve the necessary linear equations in the operator splitting
step.

\subsection{Framework}

The 3D RT equations are easiest solved in a Cartesian coordinate system
\cite[e.g., ][]{BTB97}, therefore we use a Cartesian grid of volume cells (voxels).
Basically, the voxels are allowed to have different sizes but for the tests
presented later in this paper we use fixed size voxels (as the simplest
option). The values of physical quantities, such as temperatures, opacities and
mean intensities, are averages over a voxel, which, therefore, also fixes the
local physical resolution of the grid. In the following we will specify the
size of the voxel grid by the number of voxels along each positive axis, e.g.,
$n_x = n_y = n_z = 32$ specifies a voxel grid from voxel coordinates
$(-32,-32,-32)$ to $(32,32,32)$ for a total of  $(2*32+1)^3 = 274625$ voxels,
$65$ along each axis. The voxel $(0,0,0)$ is at the center of the voxel grid.
{The voxel centers are the grid points.}
The voxel coordinates are related by grid scaling factors to physical space,
depending on the problem. The framework does not require $n_x = n_y = n_z$, we
use this for the tests presented in this paper for convenience.

The applications that we intend to solve with the 3DRT framework will 
involve optically thick environments with a significant scattering contribution,
e.g., modeling the light reflected by an extrasolar giant planet close to 
its parent star. Therefore, not only a formal solution is required but the 
full solution of the 3D radiative transfer equation with scattering. 
In this paper, we describe a method
based on the operator splitting approach. Operator splitting works best if 
a non-local $\lstar$ operator is used in the calculations \cite[e.g.,][]{aliperf},
therefore we describe a non-local $\lstar$ method here. The operator splitting
method can be combined with other methods, like multigrids, to allow for greater
flexibility and better convergence, which we will discuss in a later paper.

{
\subsection{Radiative transfer equation}

The static radiative transfer equation in 3-D may be written
\begin{equation}
\hat{\vec n} \cdot \nabla
I(\nu,\vec x,\hat{\vec n}) = \eta(\nu,\vec x) - \chi(\nu,\vec x)I(\nu,\vec
x,\hat{\vec n})
\end{equation}
where $I(\nu,\vec x,\hat{\vec n})$ is the specific intensity at frequency
$\nu$, position $\vec x$, in the direction $\hat {\vec n}$, $\eta(\nu,\vec
x)$ is the emissivity at frequency
$\nu$ and position $\vec x$, and $\chi(\nu,\vec x)$ is the total
extinction at frequency
$\nu$ and position $\vec x$. The source function $S =
{\eta}/{\chi}$. Here, we will work in the steady-state so that
${\partial I}/{\partial t} = 0$, and in Cartesian coordinates so
the $\nabla =  \frac{\partial}{\partial x} + \frac{\partial}{\partial
  y} + \frac{\partial}{\partial z}$ and the direction $\hat{\vec n}$ is
defined by two angles $(\theta,\phi)$ at the position $\vec x$.
}

{
\subsection{The operator splitting method}

The mean intensity $J$ is obtained from the source function
$S$ by a formal solution of the RTE which is symbolically written
using the $\Lambda$-operator $\Lambda$ as
\begin{equation}
     J = \Lambda S.              \label{frmsol}
\end{equation}
The source function is given by $S=(1-\e)J + \e B$, where $\e$ 
denotes the thermal coupling parameter and $B$ is Planck's function.

The $\L$-iteration method, i.e.\ to solve Eq.~\ref{frmsol} by a fixed-point
iteration scheme of the form
\bea
   \Jnew = \L \Sold , \quad
   \Snew = (1-\e)\Jnew + \e B  ,\label{alisol}
\eea
fails in the case of large optical depths and small $\e$.
{Here, $\Sold$ is the current estimate for the source
function $S$ and $\Snew$ is new, improved, extimate of $S$ for the next
iteration.} { The failure to converge of the $\Lambda$-iteration} is caused
by the fact that the largest eigenvalue of the amplification matrix
is approximately \cite[]{mkh75}
$\l_{\rm max} \approx (1-\e)(1-T^{-1})$, where $T$ is the  optical 
thickness of the medium. For small $\e$ and large $T$, this is very close
to unity and, therefore, the convergence rate of the $\L$-iteration is very 
poor. A physical description of this effect can be found in 
\cite{mih80}.

The idea of the ALI or operator splitting (OS) method is to reduce the 
eigenvalues of the amplification matrix in the iteration scheme
\cite[]{cannon73}  by 
introducing an approximate $\L$-operator (ALO) $\lstar$
and to split $\L$ according to
\begin{equation}
           \L = \lstar +(\L-\lstar) \label{alodef}
\end{equation}
and rewrite Eq.~\ref{alisol} as
\begin{equation}
     \Jnew = \lstar \Snew + (\L-\lstar)\Sold. 
\end{equation}
This relation can be written as \cite{hamann87}
\begin{equation}
    \left[1-\lstar(1-\e)\right]\Jnew = \Jfs - \lstar(1-\e)\Jold, \label{alo}
\end{equation}
where $\Jfs=\L\Sold$ { and $\Jold$ is the current
estimate of the mean intensity $J$}. Equation~\ref{alo} is solved to get the new values of 
$\Jb$ which is then used to compute the new 
source function for the next iteration cycle.

Mathematically, the OS method belongs to the same family of iterative
methods as the Jacobi or the Gauss-Seidel methods
\cite[]{golub89:_matrix}. These 
methods have the general form
\begin{equation}
    M x^{k+1} = Nx^{k} + b
\end{equation}
for the iterative solution of a linear system $Ax=b$ where the system
matrix $A$ is split according to $A=M-N$. In the case of the OS
method we have $M=1-\lstar(1-\e)$ and, accordingly,
$N=(\L-\lstar)(1-\e)$ for the system matrix $A=1-\L(1-\e)$. The
convergence of the iterations depends on the spectral radius,
$\rho(G)$, of the iteration matrix $G=M^{-1}N$.  For convergence the
condition $\rho(G)<1$ must be fulfilled, this puts a restriction on
the choice of $\lstar$. In general, the iterations will converge
faster for a smaller spectral radius.  To achieve a significant
improvement compared to the $\L$-iteration, the operator $\lstar$ is
constructed so that the eigenvalues of the iteration matrix $G$ are
much smaller than unity, resulting in swift convergence. Using
parts of the exact $\L$ matrix (e.g., its diagonal or a tri-diagonal
form) will optimally reduce the eigenvalues of the 
$G$. The
calculation and the structure of $\lstar$ should be simple in order to
make the construction of the linear system in Eq.~\ref{alo} fast. For
example, the choice $\lstar=\L$ is best in view of the
convergence rate (it is equivalent to a direct solution by matrix inversion)
but the explicit construction of $\L$ is more time
consuming than the construction of a simpler $\lstar$. The solution of
the system Eq.~\ref{alo} in terms of linear algebra, using modern
linear algebra packages such as, e.g., {\tt LAPACK} \cite[]{lapack}, is so fast that
its CPU time can be neglected for the small number of variables
encountered in 1D problems (typically the number of discrete shells
is about 50). However, for 2D or 3D problems the size of $\L$ gets
very large due to the much larger number of grid points as compared to
the 1D case.  Matrix inversions, which are necessary to solve
Eq.~\ref{alo} directly, therefore become extremely time
consuming. This makes the direct solution of Eq.~\ref{alo} more CPU intensive
even for $\lstar$'s of moderate bandwidth, except for the trivial case
of a diagonal $\lstar$.  Different methods like modified conjugate
gradient methods \cite[]{turek93} may be effective for these
2D or 3D problems.

The CPU time required for the solution of the RTE using the OS method depends
on several factors: (a) the time required for a formal solution and the
computation of $\Jfs$, (b) the time needed to construct $\lstar$, (c) the time
required for  the solution of Eq.~\ref{alo}, and (d) the number of iterations
required for convergence to the prescribed accuracy. Points (a), (b) and (c)
depend mostly on the number of spatial points, and can be assumed to be fixed
for any given configuration. However, the number of iterations required to
convergence depends strongly on the bandwidth of $\lstar$.
This indicates, that there is an {\em optimum
bandwidth} of the $\lstar$-operator which will result in the shortest possible
CPU time needed for the solution of the RTE, see \cite{aliperf}.
}

\subsection{Formal solution}

The formal solution through the voxel grid can be performed by a variety of
methods. So far, we have implemented both a short-characteristic
\cite[SC,][]{OAB} and a long-characteristic (LC) method.  { Long and
short characteristics are shown schematically in Fig.~\ref{fig:chars}. In our
current implementation, the long characteristics are followed continuously
through the voxel grid, the short characteristics start at the center
of a voxel and step closest to the center of the next voxel. The distances
along a (short or long) characteristic are then used to compute the optical
depth steps.} {Along a characteristic (either short or long), the
formal solution is computed using a piece-wise parabolic (PPM) or piece-wise
linear (PLM) interpolation and integration of the source function
\cite[]{ok87}. \cite{auer03} discusses the effect that high order interpolation
may cause problems, therefore, we automatically use piece-wise linear
interpolation if the change in the source function along the 3 points of the
PPM step would be larger than a prescribed threshold (typically factors of 100)
or if the  optical depth along the characteristic is very small (typically less
than $10^{-3}$).} Depending on the direction $(\theta,\phi)$ of the
characteristic, the formal solution proceeds through the voxel grid.

Therefore, {\em along a characteristic} [which is in the static case just a straight
line with given $(\theta,\phi)$] the transport equation is 
simply
\begin{equation}
   \div{d I}{d\tau}
 =
   I - S
\end{equation}
With this definition, 
the formal solution of the RTE along the characteristics can be written
in the following way \cite[cf.][for a derivation of 
the formulae]{ok87}
{
where we have suppressed the index labeling the characteristic; $\tau_i$ denotes the
optical depth along the characteristic with $\tau_1\equiv 0$ and $\tau_{i-1} \le \tau_i$
while $\tau$ is calculated using piecewise linear interpolation of 
$\chi$ along the characteristic, viz.\
\bea
       I(\tau_i) &=& I(\tau_{i-1}) \exp(\tau_{i-1}-\tau_i) \nonumber\\
                 & &\quad  +\int_{\tau_{i-1}}^{\tau_i} S(\tau) \exp(\tau-\tau_i)
                      \, d\tau \\
       I(\tau_i) &\equiv& I_{i-1}\exp(-\Delta\tau_{i-1})+\Delta I_i
\eea
$i$ labels the points along a characteristic { and $\Delta\tau_{i}$
is calculated using piecewise linear interpolation of $\chi$ along the
characteristic
\begin{equation}
\Delta\tau_{i-1} = (\chi_{i-1}+\chi_i)|s_{i-1}-s_i|/2
\end{equation}
The starting points
of each characteristic are 
at the center of the voxels on their starting faces of the voxel grid.}

The source function $S(\tau)$ along a characteristic is interpolated by 
linear or parabolic polynomials so that 
\begin{equation}
    \Delta I_i = \alpha_i S_{i-1}
                + \beta_i S_i + \gamma_i S_{i+1}
\end{equation}
with
\bea
   \alpha_i &=& e_{0i} 
               + [e_{2i}-(\Delta\tau_i+2\Delta\tau_{i-1})e_{1i}] \nonumber \\
            & & \qquad /[\Delta\tau_{i-1}(\Delta\tau_i+\Delta\tau_{i-1})] \\
   \beta_i  &=&  [(\Delta\tau_i+\Delta\tau_{i-1})e_{1i}-e_{2i}]
               /[\Delta\tau_{i-1}\Delta\tau_i] \\
   \gamma_i &=&  [e_{2i}-\Delta\tau_{i-1}e_{1i}]
               /[\Delta\tau_i(\Delta\tau_i+\Delta\tau_{i-1})]
\eea
for parabolic interpolation and
\bea
   \alpha_i &=& e_{0i} - e_{1i}/\Delta\tau_{i-1}\\
   \beta_i  &=& e_{1i}/\Delta\tau_{i-1}         \\
   \gamma_i &=& 0
\eea
for linear interpolation.
The auxiliary functions are given by
\bea
   e_{0i} &=& 1-\exp(-\Delta\tau_{i-1})      \\
   e_{1i} &=& \Delta\tau_{i-1} - e_{0i}      \\
   e_{2i} &=& (\Delta\tau_{i-1})^2 - 2e_{1i}
\eea
and $\Delta\tau_i \equiv \tau_{i+1}-\tau_i$ is the optical depth along the
characteristic from point $i$ to point $i+1$. The linear coefficients have to be used (at
least) at the last integration point along each characteristic, and for some 
cases it might be
better to also use linear interpolation for some inner points so as to ensure
stability.
}

The integration over solid angle can be done in the static case using a simple
Simpson or trapezoidal quadrature formula. However, in the case of Lagrangian frame
radiation transport, the angles $(\theta,\phi)$ vary along a (curved)
characteristic. Therefore, the $(\theta,\phi)$ grid changes for each voxel and
developing a quadrature formula in advance requires a pass through all voxels,
storing all $(\theta,\phi)$ points for each of them. For larger grids this will
amount to substantial long term memory requirements as the resulting quadrature
weights will have to be stored for each $(\theta,\phi)$ pair at all voxels. To
avoid this, we have implemented a simple Monte-Carlo (MC) scheme to perform the
integration over solid angle.
{ In the MC integration, the integral
$$
J = \frac{1}{4\pi} \int_0^{2\pi} \int_0^{\pi} I \sin\theta\,d\theta d\phi
$$
is replaced by a simple MC sum of the form
$$
J \approx \frac{1}{2\pi^2} \sum I \sin\theta
$$
where the sum goes over all solid angle points $(\theta,\phi)$. {
The $(\theta,\phi)$ are randomly selected and given equal weight in the MC
sums.  This also works for precribed $(\theta,\phi)$ grids as long as the
number of $(\theta,\phi)$ points is sufficiently large.} The accuracy is
improved by maintaining the normalization numerically for a unity valued test
function.  The MC method has the advantage that the solid angle points can vary
from voxel to voxel (important for configurations with velocity fields where
the transfer equation is solved in the locally co-moving frame).  }
In the static case, the accuracy of the MC method
is only insignificantly worse than that of the deterministic quadrature, 
which indicates that the MC integration will be very useful in
the case of 3D radiation transport in moving media.

\subsubsection{Computation of $\lstar$}

As demonstrated by \cite{OAB} and \cite{ok87}, the coefficients $\alpha$,
$\beta$, and $\gamma$ can be used to construct diagonal and tri-diagonal
$\lstar$ operators for 1D radiation transport problems. In fact, up to the full
$\Lambda$ matrix can be constructed by a straightforward extension of the idea
\cite[]{aliperf,s3new}. These non-local $\lstar$ operators not only lead to
excellent convergence rates but they also avoid the problem of false
convergence that is inherent in the $\Lambda$ iteration method and can also be
an issue for diagonal (purely local) $\lstar$ operators. Therefore, it is
highly desirable to implement a non-local $\lstar$ also for the 3D case.  The
tri-diagonal operator in the 1D case is simply a nearest neighbor $\lstar$ that
considers the interaction of a point with its two direct neighbors. In the 3D
case, the nearest neighbor $\lstar$ considers the interaction of a voxel with
the (up to) $3^3-1=26$ surrounding voxels (this definition considers a somewhat larger
range of voxels that a strictly face-centered view of just 6 nearest neighbors).
This means that the non-local $\lstar$ requires the storage of 27
(26 surrounding voxels plus local, i.e., diagonal effects) times the total
number of voxels $\lstar$ elements. This can be reduced, for example
if one considers only the voxels that share a face to a total of 7 elements
for each voxel. However, this would ignore the effect of characteristics
that pass 'diagonally' through a voxel and would therefore  lead to a slower
convergence rate. 

{
The construction of the $\lstar$ operator proceeds in the same way as discussed
in \cite{s3pap}. In the 3D case, the
'previous' and 'next' voxels along each characteristic must be known so that
the contributions can be attributed to the correct voxel. Therefore, we use a
data structure that attaches to each voxel its effects on its neighbors.
The scheme can be extended trivially to include longer range interactions
for better convergence rates (in particular on larger voxel grids), however
the memory requirements to simply store $\lstar$ ultimately scales like
$n^3$ where $n$ is the total number of voxels. The storage requirements can be
reduced by, e.g., using $\Lstar$'s of different widths for different voxels.
Storage requirements are not so much a problem if a domain decomposition
parallelization method is used and enough processors are available. Below we
will show some results for test cases with larger operators.

We describe here the general procedure of calculating the $\lstar$ with {\em
arbitrary} bandwidth, up to the full $\Lambda$-operator, for the method in
spherical symmetry \cite[]{aliperf}. The construction of the $\lstar$ is
described in \cite{ok87}, so that we here summarize the relevant formulae.  In the
method of \cite{ok87}, the elements of the row of $\lstar$ are computed by setting the
incident intensities (boundary conditions) to zero and setting
$S(i_x,i_y,i_z)=1$ for one voxel $(i_x,i_y,i_z)$ and performing a formal
solution analytically.

We describe the  construction of $\lstar$ using the example of a single
characteristic. The contributions to the $\Lstar$ at a voxel $j$ are given by
\vbox{\bea
     \Lambda_{i,j} = 0  &\quad {\rm for\ }  i<j-1  \\
     \Lambda_{j-1,j} = \gamma_{j-1}   &\quad {\rm for\ }  i=j-1 \\
     \Lambda_{j,j} = \Lambda_{j-1,j}\exp(-\Delta\tau_{j-1}) +
\beta^k_{j}  
                                        &\quad{\rm for\ }  i=j  \\
     \Lambda_{j+1,j} = \Lambda_{j,j}\exp(-\Delta\tau_{j}) + 
\alpha_{j+1} 
                                        &\quad{\rm for\ }  i=j+1  \\
     \Lambda_{i,j} = \Lambda_{i-1,j}\exp(-\Delta\tau_{i-1}) 
                                        &\quad{\rm for\ }  j+1 < i \label{wide}
\eea}
These contributions are computed along a characteristic, here $i$ labels the
voxels {\em along} the characteristic under consideration. These contributions
are integrated over solid angle with the same method (either deterministic or
through the Monte-Carlo integration) that is used for the computation of the
$J$. For a nearest neighbor $\lstar$, the process of Eq.~\ref{wide} is stopped
with $i=j+1$, otherwise it is continued until the required bandwidth has been
reached (or the characteristic has reached an outermost voxel and terminates).
}

\subsection{Operator splitting step}

For a diagonal $\lstar$ the solution of Eq.~\ref{alo} is just a division by
$\left[1-\lstar(1-\e)\right]$ which requires very little effort. For the
non-local $\lstar$ operator, a matrix equation has to be solved at each step in
order to compute $\Jnew$.  For the nearest neighbor operator the matrix
structure is that of a sparse band matrix where the bandwidth of the matrix is
proportional to the square of the maximum number of points along a coordinate
and there are a total of 27 non-zero entries in every column of the matrix.
This system can be solved by any suitable method. For example, we have
implemented a solver based on {\tt LAPACK} \cite[]{lapack} routines. Although this solver works
fine for small grids [$(2*16+1)^3$ voxels with a bandwidth of 1123], the memory
requirements rise beyond available single-CPU limits with already only $(2*32+1)^3$ voxels
with a bandwidth of 4291. For a more realistic grid of $(2*256+1)^3$ voxels the
bandwidth is more than 250,000.  Therefore, a standard band matrix solver is
only useful for comparison and testing on very small grids. Other sparse 
matrix solvers \cite[for example,][]{super_lu03} have similar memory 
scaling properties.

As an alternative to direct solutions of Eq.~\ref{alo} we have implemented
iterative solvers. These solvers have the huge advantage that their
memory requirements are minimal to modest. In addition, they can be
tailored to the special format of the matrix in Eq.~\ref{alo} which makes 
coding the methods very simple.
We obtained extremely good results with 
the Jordan and the Gauss-Seidel methods, which converge rapidly to a
relative accuracy of $10^{-10}$. For either method, Ng acceleration
turned out to be very useful, reducing the number of iterations significantly.
 In addition, the Rapido method \cite[]{matrizen} which was constructed to solve
systems of the form 
\begin{equation}
(1-M)x = a
\end{equation}
can be used. However, Rapido has strong constraints on the spectral
radius of $M$ and this cannot be used for problems that involve 
substantial scattering. In the test calculations discussed below we 
have used the Jordan or Gauss-Seidel solvers as they are the fastest
of the solvers we have implemented. We verified that all solvers 
give the same results. 

\section{Application examples}

As a first step we have implemented the method as a MPI parallelized
Fortran 95 program. The parallelization of the formal solution is presently implemented
over solid angle space as this is the simplest parallelization 
option and also one of the most efficient (a domain decomposition parallelization
method will be discussed in a subsequent paper). In addition, the 
Jordan solver of the Operator splitting equations is parallelized with 
MPI (see below for scaling properties of the MPI implementation). The number
of parallelization related statements in the code is small, about 320 out
of a total of about 7900.

Our basic continuum scattering test problem is similar to that discussed in
\citet{s3pap} and in \citet{s3new}. { This test problem covers a
large dynamic range of about 9 dex in the opacities and overall optical depth
steps along the characteristics and, in our experience, constitutes a
reasonably challenging setup for the radiative transfer code.{The
application of the 3D code to 'real' problems is in preparation and requires a
substantial amount of development work (in progress). For the 1D code we have
found that the test case is actually pretty much a worst case scenario and that
it generally works better in real world problems. }} We use a sphere with a
grey continuum opacity parameterized by a power law in the continuum optical
depth $\tstd$. The basic model parameters are
\begin{enumerate}
\item Inner radius $r_c=10^{13}\,$cm, outer radius $\rout = 1.01\alog{15}\,$cm.
\item Minimum optical depth in the continuum $\t_{\rm std}^{\rm min} =
10^{-4}$ and maximum optical depth in the continuum $\t_{\rm std}^{\rm
max} = 10^{4}$.
\item Grey temperature structure with $\Teff=10^4$~K.
\item Outer boundary condition $I_{\rm bc}^{-} \equiv 0$ and diffusion
inner boundary condition for all wavelengths.
\item Continuum extinction $\chi_c = C/r^2$, with the constant $C$
fixed by the radius and optical depth grids.
\item Parameterized coherent \& isotropic continuum scattering by
defining
\begin{equation}
\chi_c = \epsilon_c \kappa_c + (1-\epsilon_c) \sigma_c
\end{equation}
with $0\le \epsilon_c \le 1$. 
$\kappa_c$ and $\sigma_c$ are the
continuum absorption and scattering coefficients.
\end{enumerate}
{ The test model is just 
an optically thick sphere put into the 3D grid.}
This problem is used because the results can be directly compared
with the results obtained with { our} 1D spherical radiation transport
code \cite[]{s3pap} to assess the accuracy of the method.
The sphere is { centered} at the center of the  Cartesian grid, which is 
in each axis 10\% larger than the radius of the sphere.
For the test calculations we use voxel grids with the same
number of spatial points in each direction (see below). The
solid angle space was discretized in $(\theta,\phi)$ with 
$n_\theta=n_\phi$ if not stated otherwise. In the following 
we discuss the results of various tests. In all tests we use
the LC method for the 3D RT solution unless stated otherwise.

\subsection{LTE tests}

In this test we have set $\epsilon=1$ to test the accuracy of the formal
solution by comparing to the results of the 1D code.  The 1D solver uses 50
radial points, distributed logarithmically in optical depth. {For the
3D solver we tested `small' grids with $n_x=n_y=n_z=2*32+1$ points along each
axis, for a total of $65^3 \approx 2.7\alog{5}$ voxels, as well  as `medium'
($n_x=n_y=n_z=2*64+1$ with a total of $129^3\approx 2.1\alog{6}$ voxels) and
'large ($n_x=n_y=n_z=2*96+1$ with a total of $193^3\approx 7.2\alog{6}$
voxels). The large grid was limited by available memory for the storage of the
non-local $\Lstar$ operator.} The solid angle space discretization uses,
{ in general,} $n_\theta=n_\phi=64$ points. In
Fig.~\ref{fig:LTE:distance} we show the mean intensities as function of
distance from the center for both the 1D ($+$ symbols) and the 3D solver. The
results show excellent agreement between the two solutions, thus the
3D RT formal solution is comparable in accuracy with the 1D formal solution.
We demonstrate below that for the conditions used in these tests a larger
number of solid angle points significantly improves the accuracy of the mean
intensities. 

\subsection{Tests with continuum scattering}
\subsubsection{$\epsilon=10^{-4}$}

For this test, we use the same basic structure setup as for the LTE test,
however we now use $\epsilon = 10^{-4}$ for a scattering dominated atmosphere.
The comparison with the results obtained with the 1D RT code (with 100 radial
points) are compared to the 3D RT code results in
Fig.~\ref{fig:eps=-4:distance}.  Note the factor of about 1000 difference
between the solution for $\epsilon = 10^{-4}$ and the results of the formal
solution with $S=B$ shown in Fig.~\ref{fig:LTE:distance} at the largest
distances (the iterations were started with $S=B$).
{Figure~\ref{fig:eps=-4:distance} shows the results for the
  `large' and `medium' spatial grids 
and for 2 different solid angle discretizations ($64^2$ and $16^2$
solid angle points).} For both the 1D and the
3D code the mean intensities were iterated to a relative accuracy of $10^{-8}$
(see below for details on convergence rates). The graph highlights the need for
a rather fine solid angle grid for the test problem. With a small number of
angular points (bottom panel in Fig.~\ref{fig:eps=-4:distance}) the numerically
induced spread of the mean intensities is significantly larger than with a
$4^2$ finer angular grid (shown in the middle panel).  {A change in spatial grid
resolution has a much smaller effect on the results than the 
number of solid angle points as shown in the top two panels
of Fig.~\ref{fig:eps=-4:distance}. The spatial resolution of the  `small' grid is, however,
too coarse to represent the test problem, in particular in the inner 
parts of the structure.}

The importance of the angular resolution is also demonstrated in 
Fig.~\ref{fig:eps=-4:contour} which shows the contours of the mean 
intensity on the six `faces' of the voxel cube for $129^3$ spatial points
and $16^2$ (left panel) and $64^2$ (right panel) solid angles. The calculation
with $64^2$ solid angle points shows symmetric contours whereas the 
smaller $16^2$ angle points model shows asymmetries and `banding' like
structures on all faces (in particular on the $x-y$ faces). This clearly
indicates that for the test conditions used the angular resolution has
to be larger than $16^2$ for accurate solutions. This can also be seen
in surface plots of the mean intensities at the $-n_{\rm x}$ faces of the
$z-y$ planes for both calculations, cf. Fig.~\ref{fig:eps=-4:surface}.
The surface calculated with $64^2$ angles is much smoother and shows 
no or little banding compared to the surface for $16^2$ angular points.
The effects of higher spatial resolution on the $J$ surface at the 
$z-y$ face is shown in Fig.~\ref{fig:eps=-4:surface96}. Clearly, $64^2$ 
angles produce a smooth surface without significant artifacts for the
test case.

The convergence properties of the various methods for this test case are shown
in Fig.~\ref{fig:eps=-4:iteration}. As expected, the $\Lambda$ iteration
converges very slowly and requires more than 300 iterations to reach a relative
error of less than $10^{-8}$. 
The diagonal $\Lstar$ operator converges
significantly  better than the $\Lambda$ iteration but is still rather slowly
convergent. The nearest neighbor $\Lstar$ converges substantially faster
(by more than a factor of 3 than the diagonal $\Lstar$).
The diagonal and nearest neighbor iterations can be accelerated by using Ng's
method \cite[]{ng74}, as shown in Fig.~\ref{fig:eps=-4:iteration}. In both cases, a 4th
order Ng acceleration was used after 20 initial iterations, starting Ng
acceleration too early can lead to convergence failures since Ng acceleration
is based on extrapolation. { An attempt to 
apply the Ng acceleration to the $\Lambda$ iteration was not successful, 
similar to the 1D case, as the convergence rate of the $\Lambda$ iteration 
appears to be too small to be useful for the Ng method.} For comparison, we show
in Fig.~\ref{fig:eps=-4:iteration} the convergence properties of the 1D RT
solver with a tri-diagonal $\Lstar$ and Ng acceleration (here started much
earlier). The convergence rates of the 1D tri-diagonal and 3D nearest neighbor
methods are very comparable. The convergence properties are also relevant for
the overall speed of the method: whereas the time for the formal solution (for
a given number of voxels and processors) depends roughly linearly on the total
number of solid angle points, the time for the solution of the operator
splitting step does not depend on the number of angle points and actually 
{\em decreases} with iteration number if the Jordan, Gauss-Seidel or 
Rapido solvers are used. Therefore, the nearest neighbor operator will
become more and more efficient as the number of angles and/or voxels
increases. 

{The smaller initial corrections of the $\Lambda$ 
iteration are actually an indication that it just corrects too little,
the operator splitting method gives initially much larger corrections. 
This is exactly like
similar test cases in 1D that we have tried. 
The initial corrections are so large
because the initial guesses are very wrong (intentionally) so that between
initial guess and final result we have changes by close to 10 orders
of magnitude. This means that the initial corrections of the 
operator splitting method have to be large.}

The convergence rates will depend on the grid resolution as 
the optical depths between voxels will be smaller for larger
resolutions and thus the coupling between voxels will be 
stronger. This effect is shown in Fig.~\ref{fig:eps=-4:iteration:grids}
where we show the convergence rates for various grids. The convergence
rates are independent of the number of solid angle points 
but depend weakly on the number of grid points, as expected. Thus,
for voxel grids with higher resolution, a larger $\Lstar$ (more neighbors)
will be more useful than for coarser voxel grids.

Figure~\ref{fig:wide} shows the convergence rates for the $\epsilon=10^{-4}$
test case for different $\Lstar$ bandwidths.  The wider $\Lstar$'s lead to
improved convergence rates; however, the net wallclock time increases for the
wider $\Lstar$'s in the parallel code. This is caused by the substantially
larger amount of information that needs to be passed between processes in order
to build the wider $\Lstar$. In addition, the memory requirements for a {\em
given} number of voxels scales like the cube of the number of neighbors
considered in the $\Lstar$. As Fig.~\ref{fig:wide} demonstrates, the
convergence is strongly improved by the use of Ng acceleration, however, for
larger $\Lstar$'s the Ng method can be detrimental for wider $\Lstar$ compared
to narrower $\Lstar$.

\subsubsection{$\epsilon=10^{-8}$}

The final test we present in this paper considers a case with a 
much larger scattering contribution: $\epsilon = 10^{-8}$. The results
for a grid with  $129^3$ voxels and $64^2$ angular points are shown in 
Fig.~\ref{fig:eps=-8:distance}. Fig.~\ref{fig:eps=-8:axes} shows the results
for slices along the coordinate axes (the two coordinates being centered, respectively).
Even though the dynamic range of the mean intensities is huge, nearly
12 dex, the 
results are quite accurate.
For this case the lack of spatial resolution in the inner
parts of the voxel grid can be seen. Here $|\nabla J|$ is huge and cannot
be fully resolved by the 3D RT code (the 1D code naturally has much higher 
resolution). However, only a few voxels away from the center 
the results agree very well.

The convergence
plots in Fig~\ref{fig:eps=-8:iteration} show the results for a very difficult
test case with $\t_{\rm std}^{\rm max} = 10^{8}$ for a fixed voxel 
and solid angle grid with $65^3$ voxels and $16^2$ angular points. 
For this test, the $\Lambda$ iteration fails completely. The diagonal
operator provides significant speed-up, but still requires
more than 1600 iterations to reach the required convergence limit.
For this test, the Ng acceleration does not work with the diagonal operator,
the iteration process failed immediately after it was started. It is likely that
a better result could be obtained if Ng acceleration is started in the 
steep part of the diagonal operator's convergence, e.g., after about 500 
regular iterations. The nearest neighbor operator leads to much faster
convergence, even without Ng acceleration the solution converges in about 
450 iterations. Here, Ng acceleration works very well with the nearest
neighbor operator, convergence is reached after 177 iterations. This is 
still about a factor of 2 more than for the 1D code, but much better
than with the diagonal operator. The variation of the convergence
rate for the nearest neighbor operator and Ng acceleration with 
the size of the solid angle grid is shown 
in Fig.~\ref{fig:eps=-8:iteration:grids}. The case with the smallest
angular grid ($16^2$ points) actually converges \emph{more slowly} than the 
$32^2$ and $64^2$ grids. The higher resolution grids convergence rate compares well with the 
1D code. This highlights the importance of the non-local $\Lstar$ operator
and a large enough solid angle grid for rapid convergence and accuracy.

\subsubsection{Visualization}

We have implemented a simple visualization of the results in order
to view images of the emitted intensities. The visualization uses IDL
to read the results of the 3D RT code, performs a formal
solution for a specific $(\theta,\phi)$  and displays the result
as an image of the intensities leaving the voxel grid.
{ These figures are extremely helpful for discovering even 
small problems with the generation and handling of the characteristics
or issues due to low resolution in solid angle or space.
Such problems will immediately show up as asymmetries in the 
generated images.}
We show 
the results for the LTE test (with $65^3$ voxels) in 
Fig.~\ref{fig:vis:LTE}. The limb  darkening of the test sphere
is clearly visible in the figures. The slight pixellation and asymmetries
are due to spatial and angular resolution. Note that the scales of the different
panels are different due to the changing orientation of the data
cube. The generated images for the $\epsilon=10^{-4}$ test case 
with $129^3$ and $193^3$ voxels are shown in Figs.~\ref{fig:vis:eps=-4:64}
and \ref{fig:vis:eps=-4:96}, respectively. For both figures, the
source functions were calculated with $64^2$ angles. The images show 
much less pixellation and far less artifacts than the images shown
in Fig.~\ref{fig:vis:LTE}. The results for the $\epsilon=10^{-8}$ test
case are shown in Fig.~\ref{fig:vis:eps=-8:64}, they look similar to
the $\epsilon=10^{-4}$ case with the same grid sizes.

\subsubsection{MPI scaling properties}

Figure~\ref{fig:scaling} shows the scaling properties of the 
MPI version of the 3D radiation transport code for the $\epsilon=10^{-8}$
test case. The runs were performed on 2 parallel compute clusters,
one equipped with 1.8GHz dual Opteron CPUs and an Infiniband interconnect from
Delta computer and one equipped with 2.0GHz dual G5 CPUs with Gbit ethernet
network from Apple computer (Xserves). The speedup we obtain in
the MPI version is close to optimal, about a factor of 28 with 32 MPI processes
on 32 CPUs (or 16 compute nodes). The fact that the speedup is very good
shows that the load balancing is optimal and that the time spent in the 
MPI communication routines is negligible compared to the compute times. 

With $129^3$ voxels, the code uses about 0.6GB of memory. With 10 CPUs and
$64^2$ angles, the wallclock time for a formal solution is about 310\,sec
(400\,sec) on 2.0GHz Xserve G5s (on 1.8GHz Opterons), 9\,sec (9\,sec) for the
required  MPI communication, and between 3--26\,sec (12--120\,sec) to solve the
linear system. Since the linear system is solved iteratively, the time for the
solution is reduces as the overall convergence limit is approached.

\section{Conclusions}

We have described a framework for solving three-dimensional radiative transfer
problems in scattering dominated environments.  The method uses a non-local
operator splitting technique to solve the scattering problem. The formal
solution is based on a long characteristic piece-wise parabolic procedure. For
strongly scattering dominated test cases (sphere in a box) we find good
convergence with non-local $\Lstar$ operators, as well as minimal numerical
diffusion with the long characteristics method and adequate resolution. A
simple MPI parallelization gives excellent speedups on parallel clusters.
In subsequent work we will implement a domain decomposition method to 
allow much larger spatial grids. { Presently, we have implemented
the method for static media, it can be used without significant changes to 
solve problems in the Eulerian-frame for media with low velocity fields.}
{The distribution of matter over the voxels is,
in the general 3D case, arbitrary. We chose a spherical test case
to be able to compare the results our 1D code.}

In Figs.~\ref{fig:eps=-4:contour:SC} and \ref{fig:eps=-4:surface:SC} we compare
the results for the $\epsilon=10^{-4}$ test case using a simple implementation
of the  short characteristics method and the long characteristics method used
in this paper. The test grid contains $129^3$ voxel and $64^2$ angular points.
The high diffusivity of the SC method is evident. Other authors
\cite[]{steiner91,BTB97,TBB95,vath94,ABTB94,vannoort02} have used SC methods in
multi-dimensional radiative transport problems.  Short characteristics
techniques are faster but require special considerations to reduce numerical
diffusion \cite[]{auer03}. 
{ We may further look into the SC method in later papers.}

We have generalized the operator splitting to include larger bandwidth
operators. They lead to faster convergence although they do require more memory
and ultimately more computing time. Nevertheless, they will be useful for
highly complex problems and we have developed a highly flexible approach to the
construction of the $\Lstar$ operator so that the bandwidth may be set for each
spatial point individually as the problem and computational resources require.

We have designed an especially general and flexible framework for 3D 
radiative transfer problems with scattering. In future papers of 
this series we will describe its extension to line transfer problems,
multi-level NLTE calculations, and differentially moving flows.

\begin{acknowledgements}
This work was supported in part by by NASA grants NAG5-12127 and NAG5-3505, NSF
grants AST-0204771 and AST-0307323. PHH was supported in part by the P\^ole
Scientifique de Mod\'elisation Num\'erique at ENS-Lyon. Some of the
calculations presented here were performed at the National Energy Research
Supercomputer Center (NERSC), which is supported by the Office of Science of
the U.S.  Department of Energy under Contract No. DE-AC03-76SF00098; at the
H\"ochstleistungs Rechenzentrum Nord (HLRN); and at the Hamburger Sternwarte
Apple G5 and Delta Opteron clusters financially supported by the DFG and the
State of Hamburg.  We thank all these institutions for a generous allocation of
computer time.
\end{acknowledgements}


\bibliography{yeti,radtran,rte_yeti}

\begin{thebibliography}{31}
\expandafter\ifx\csname natexlab\endcsname\relax\def\natexlab#1{#1}\fi

\bibitem[{{Adam}(1990)}]{adam90}
{Adam}, J. 1990, A\&A, 240, 541

\bibitem[{Anderson {et~al.}(1992)Anderson, Bai, Bischof, Demmel, Dongarra,
  Croz, Greenbaum, Hammarling, McKenney, Ostrouchov, \& Sorensen}]{lapack}
Anderson, E., Bai, Z., Bischof, C., {et~al.} 1992, LAPACK Users' Guide (SIAM)

\bibitem[{Auer(2003)}]{auer03}
Auer, L. 2003, in Stellar Atmosphere Modeling, ed. I.~Hubeny, D.~Mihalas, \&
  K.~Werner, Vol. 288 (San Franscisco: ASP Conf. Series), 3

\bibitem[{{Auer} {et~al.}(1994){Auer}, {Fabiani Bendicho}, \& {Trujillo
  Bueno}}]{ABTB94}
{Auer}, L., {Fabiani Bendicho}, P., \& {Trujillo Bueno}, J. 1994, A\&A, 292,
  599

\bibitem[{Cannon(1973)}]{cannon73}
Cannon, C.~J. 1973, JQSRT, 13, 627

\bibitem[{{Dykema} {et~al.}(1996){Dykema}, {Klein}, \& {Castor}}]{DKC96}
{Dykema}, P.~G., {Klein}, R.~I., \& {Castor}, J.~I. 1996, ApJ, 457, 892

\bibitem[{{Fabiani Bendicho} {et~al.}(1997){Fabiani Bendicho}, {Trujillo
  Bueno}, \& {Auer}}]{BTB97}
{Fabiani Bendicho}, P., {Trujillo Bueno}, J., \& {Auer}, L. 1997, A\&A, 324,
  161

\bibitem[{Golub \& {Van Loan}(1989)}]{golub89:_matrix}
Golub, G.~H. \& {Van Loan}, C.~F. 1989, Matrix computations (Baltimore: Johns
  Hopkins University Press)

\bibitem[{Hamann(1987)}]{hamann87}
Hamann, W.-R. 1987, in Numerical Radiative Transfer, ed. W.~Kalkofen (Cambridge
  University Press), 35

\bibitem[{Hauschildt(1992)}]{s3pap}
Hauschildt, P.~H. 1992, JQSRT, 47, 433

\bibitem[{{Hauschildt} \& {Baron}(2004)}]{s3new}
{Hauschildt}, P.~H. \& {Baron}, E. 2004, \aap, 417, 317

\bibitem[{Hauschildt {et~al.}(1994)Hauschildt, St{\"o}rzer, \& Baron}]{aliperf}
Hauschildt, P.~H., St{\"o}rzer, H., \& Baron, E. 1994, JQSRT, 51, 875

\bibitem[{{Hummel}(1994{\natexlab{a}})}]{hummel94a}
{Hummel}, W. 1994{\natexlab{a}}, Astrophys. Space. Sci., 216, 87

\bibitem[{{Hummel}(1994{\natexlab{b}})}]{hummel94b}
{Hummel}, W. 1994{\natexlab{b}}, A\&A, 289, 458

\bibitem[{{Hummel} \& {Dachs}(1992)}]{hummeldachs92}
{Hummel}, W. \& {Dachs}, J. 1992, A\&A, 262, L17

\bibitem[{Mihalas(1978)}]{mihalas78sa}
Mihalas, D. 1978, Stellar Atmospheres (New York: W. H. Freeman)

\bibitem[{Mihalas(1980)}]{mih80}
Mihalas, D. 1980, ApJ, 237, 574

\bibitem[{Mihalas {et~al.}(1975)Mihalas, Kunasz, \& Hummer}]{mkh75}
Mihalas, D., Kunasz, P., \& Hummer, D. 1975, ApJ, 202, 465

\bibitem[{Ng(1974)}]{ng74}
Ng, K.~C. 1974, J. Chem. Phys., 61, 2680

\bibitem[{{Norman}(2000)}]{ZEUS3D}
{Norman}, M.~L. 2000, in Revista Mexicana de Astronomia y Astrofisica
  Conference Series, Vol.~9, 66--71

\bibitem[{Olson {et~al.}(1987)Olson, Auer, \& Buchler}]{OAB}
Olson, G.~L., Auer, L.~H., \& Buchler, J.~R. 1987, JQSRT, 38, 431

\bibitem[{Olson \& Kunasz(1987)}]{ok87}
Olson, G.~L. \& Kunasz, P.~B. 1987, JQSRT, 38, 325

\bibitem[{{Papkalla}(1995)}]{papkalla95}
{Papkalla}, R. 1995, A\&A, 295, 551

\bibitem[{Rijkhorst {et~al.}(2005)Rijkhorst, Plewa, Dubey, \& Mellema}]{RPDM05}
Rijkhorst, E.-J., Plewa, T., Dubey, A., \& Mellema, G. 2005, A\&A, in press,
  astro-ph/0505213

\bibitem[{{Steiner}(1991)}]{steiner91}
{Steiner}, O. 1991, A\&A, 242, 290

\bibitem[{{Trujillo Bueno} \& {Fabiani Bendicho}(1995)}]{TBB95}
{Trujillo Bueno}, J. \& {Fabiani Bendicho}, P. 1995, ApJ, 455, 646

\bibitem[{Turek(1993)}]{turek93}
Turek, S. 1993, preprint

\bibitem[{{van Noort} {et~al.}(2002){van Noort}, Hubeny, \& Lanz}]{vannoort02}
{van Noort}, M., Hubeny, I., \& Lanz, T. 2002, ApJ, 568, 1066

\bibitem[{{Vath}(1994)}]{vath94}
{Vath}, H.~M. 1994, A\&A, 284, 319

\bibitem[{Xiaoye \& Demmel(2003)}]{super_lu03}
Xiaoye, S.~L. \& Demmel, J.~W. 2003, ACM Transaction on Mathematical Software,
  29, 110

\bibitem[{Zurm{\"u}hl \& Falk(1986)}]{matrizen}
Zurm{\"u}hl, R. \& Falk, S. 1986, Matrizen und ihre Anwendungen, 5th edn.,
  Vol.~2 (Berlin: Springer-Verlag)

\end{thebibliography}

\begin{figure}
\centering
\begin{minipage}{0.4\hsize}
\end{minipage}
\begin{minipage}{0.4\hsize}
\end{minipage}
\caption{\label{fig:chars}{ Schematic sketch of the different
types of characteristics used in the framework. The left panel 
shows the long characteristics, the right panel the short characteristics.
The voxel boundaries and centers ('$+$' symbols) are indicated.
The '$+$' denote the points between which the geometric distance 
is used to compute optical depth steps.}}
\end{figure}

\begin{figure}
\centering
\caption{\label{fig:LTE:distance} Comparison of the results obtained 
for the LTE test with the 1D solver ($+$ symbols) and the 3D solver.
The $x$ axis shows the distances from the center of the sphere, the 
$y$ axis the $\log$ of the  mean intensity.}
\end{figure}

\begin{figure}
\centering
\caption{\label{fig:eps=-4:distance} Comparison of the results obtained 
for the scattering dominated ($\epsilon=10^{-4}$) test with the 1D solver ($+$ symbols) and the 3D solver.
The $x$ axis shows the distances from the center of the sphere, the 
$y$ axis the $\log$ of the  mean intensity. The top panel shows the 
results for a (spatial; solid angle) grid with $(193^3; 64^2)$ points, the middle panel 
for $(129^3; 64^2)$ points and the bottom panel for $(129^3; 16^2)$ points.
}
\end{figure}

\begin{figure*}
\centering
\begin{minipage}{0.4\hsize}
\end{minipage}
\begin{minipage}{0.4\hsize}
\end{minipage}
\caption{\label{fig:eps=-4:contour} Mean intensity contour plots for the test case 
with $\epsilon=10^{-4}$ with $129^3$ spatial points and
$16^2$ (left panel) and $64^2$ solid angle points. The axes are labeled
by voxel index with $(x,y,z) = (0,0,0)$ being the center of the 
voxel grid. Each plot shows one outside face of the voxel cube 
(the physical scales are the same in all directions).}
\end{figure*}

\begin{figure}
\centering
\begin{minipage}{0.4\hsize}
\end{minipage}
\begin{minipage}{0.4\hsize}
\end{minipage}
\caption{\label{fig:eps=-4:surface} Mean intensity surfaces at the 
$z-y$ face for the test case 
with $\epsilon=10^{-4}$, $129^3$ spatial points, and
$16^2$ (left panel) and $64^2$ solid angle points. The axes are labeled
by voxel index with $(x,y,z) = (0,0,0)$ being the center of the 
voxel grid. Each plot shows one outside face of the voxel cube 
(the physical scales are the same in all directions).}
\end{figure}

\begin{figure}
\centering
\caption{\label{fig:eps=-4:surface96} Mean intensity surface at the 
$z-y$ face for the test case 
with $\epsilon=10^{-4}$ with $193^3$ spatial points and
$64^2$ solid angle points. The axes are labeled
by voxel index with $(x,y,z) = (0,0,0)$ being the center of the 
voxel grid.}
\end{figure}

\begin{figure}
\centering
\caption{\label{fig:eps=-4:iteration} Convergence properties
of the codes for the $\epsilon=10^{-4}$ test case. The labels 
indicate the different methods used. The 3D test runs use
$65^3$ spatial and $16^2$ angular points.}
\end{figure}

\begin{figure}
\centering
\caption{\label{fig:eps=-4:iteration:grids} Convergence properties
of the codes for the $\epsilon=10^{-4}$ test case and different grid sizes. The labels 
indicate the different grid sizes used, all but the $\Lambda$ iteration use
the nearest neighbor operator with Ng acceleration.}
\end{figure}

\begin{figure}
\centering
\caption{\label{fig:wide} Convergence properties of the $\epsilon=10^{-4}$ 
test case for various $\Lstar$ operator bandwidth choices with and without
Ng acceleration.}
\end{figure}

\begin{figure}
\centering
\caption{\label{fig:eps=-8:distance} Comparison of the results obtained 
for the scattering dominated ($\epsilon=10^{-8}$) test with the 1D solver ($+$ symbols) and the 3D solver.
The $x$ axis shows the distances from the center of the sphere, the 
$y$ axis the $\log$ of the  mean intensity. The graph shows the 
results for a (spatial; solid angle) grid with $(129^3; 64^2)$ points. Note
the large dynamic range (12 dex) of the mean intensities.
}
\end{figure}

\begin{figure}
\centering
\caption{\label{fig:eps=-8:axes} Comparison of the results obtained 
for the scattering dominated ($\epsilon=10^{-8}$) test with the 1D solver ($+$ symbols) and the 3D solver
for slices along the $x$, $y$, and $z$ axes. 
The plot shows the 
results for a (spatial; solid angle) grid with $(129^3; 64^2)$ points. Note
the large dynamic range (12 dex) of the mean intensities.
}
\end{figure}

\begin{figure}
\centering
\caption{\label{fig:eps=-8:iteration} Convergence properties
of the codes for the $\epsilon=10^{-8}$ test case. The labels 
indicate the different methods used. These test runs use
$65^3$ spatial and $16^2$ angular points.}
\end{figure}

\begin{figure}
\centering
\caption{\label{fig:eps=-8:iteration:grids} Convergence properties
of the codes for the $\epsilon=10^{-8}$ test case and different angular grid sizes. The labels 
indicate the different grid sizes used, all but the $\Lambda$ iteration use
the nearest neighbor operator with Ng acceleration.}
\end{figure}

\begin{figure}
\centering
\begin{minipage}{0.4\hsize}
\end{minipage}
\begin{minipage}{0.4\hsize}
\end{minipage}
\caption{\label{fig:vis:LTE} Visualization of the results for the LTE case.
The voxel grid has $65^3$ elements. The intensity image is shown for
$(\theta,\phi) = (0,0)$ (upper left panel), $(45,45)$ (bottom right panel),
$(140,250)$ (upper right panel), and $(89,139)$ (bottom right panel) degrees.
The intensities are mapped linearly to 255 shades of gray. The direction of the
axes are given with axes lengths corresponding to a total of 50 voxels. The
borders of the front faces of the voxel cube are shown, their widths corresponds to
the apparent width of a voxel. Note that
the different panels are at different scales.}
\end{figure}

\begin{figure}
\centering
\begin{minipage}{0.4\hsize}
\end{minipage}
\begin{minipage}{0.4\hsize}
\end{minipage}
\caption{\label{fig:vis:eps=-4:64} Visualization of the results for the
$\epsilon=10^{-4}$ case with a $129^3$ elements voxel grid.  The intensity image
is shown for $(\theta,\phi) = (0,0)$ (upper left panel), $(45,45)$ (bottom
right panel), $(140,250)$ (upper right panel), and $(89,139)$ (bottom right
panel) degrees.  The intensities are mapped linearly to 255 shades of gray. The
direction of the axes are given with axes lengths corresponding to a total of
100 voxels. The
borders of the front faces of the voxel cube are shown, their widths corresponds to
the apparent width of a voxel. Note that the different panels are at different scales.}
\end{figure}

\begin{figure}
\centering
\begin{minipage}{0.4\hsize}
\end{minipage}
\begin{minipage}{0.4\hsize}
\end{minipage}
\caption{\label{fig:vis:eps=-4:96} Visualization of the results for the
$\epsilon=10^{-4}$ case with a $193^3$ voxel grid.  The intensity image
is shown for $(\theta,\phi) = (0,0)$ (upper left panel), $(45,45)$ (bottom
right panel), $(140,250)$ (upper right panel), and $(89,139)$ (bottom right
panel) degrees.  The intensities are mapped linearly to 255 shades of gray. The
direction of the axes are given with axes lengths corresponding to a total of
100 voxels. The
borders of the front faces of the voxel cube are shown, their widths corresponds to
the apparent width of a voxel. Note that the different panels are at different scales.}
\end{figure}

\begin{figure}
\centering
\begin{minipage}{0.4\hsize}
\end{minipage}
\begin{minipage}{0.4\hsize}
\end{minipage}
\caption{\label{fig:vis:eps=-8:64} Visualization of the results for the
$\epsilon=10^{-8}$ case with a $129^3$ elements voxel grid.  The intensity image
is shown for $(\theta,\phi) = (0,0)$ (upper left panel), $(45,45)$ (bottom
right panel), $(140,250)$ (upper right panel), and $(89,139)$ (bottom right
panel) degrees.  The intensities are mapped linearly to 255 shades of gray. The
direction of the axes are given with axes lengths corresponding to a total of
100 voxels. The
borders of the front faces of the voxel cube are shown, their widths corresponds to
the apparent width of a voxel. Note that the different panels are at different scales.}
\end{figure}
\clearpage

\begin{figure}
\centering
\caption{\label{fig:scaling} Scaling properties of the MPI version
of the 3D RT code for parallel clusters based on Opterons and G5 CPUs.
In absolute scales the G5s are about 30\% faster than the Opterons.}
\end{figure}

\begin{figure*}
\centering
\begin{minipage}{0.4\hsize}
\end{minipage}
\begin{minipage}{0.4\hsize}
\end{minipage}
\caption{\label{fig:eps=-4:contour:SC} Comparison of the mean intensity contour
plots for the test case with $\epsilon=10^{-4}$ with $129^3$ spatial points and
$64^2$ solid angle points. The left panel shows the results obtained with the 
short characteristics method whereas the right panel shows the results of
the long characteristics method. The axes are labeled by
voxel index with $(x,y,z) = (0,0,0)$ being the center of the voxel grid. Each
plot shows one outside face of the voxel cube (the physical scales are the same
an all directions).}
\end{figure*}

\begin{figure}
\centering
\begin{minipage}{0.4\hsize}
\end{minipage}
\begin{minipage}{0.4\hsize}
\end{minipage}
\caption{\label{fig:eps=-4:surface:SC} Comparison of the mean intensity surfaces at the 
$z-y$ face for the test case 
with $\epsilon=10^{-4}$ with $129^3$ spatial points and
$64^2$ solid angle points. The left panel shows the results obtained with the 
short characteristics methods whereas the right panel shows the results of
the long characteristics method.  The axes are labeled
by voxel index with $(x,y,z) = (0,0,0)$ being the center of the 
voxel grid. Each plot shows one outside face of the voxel cube 
(the physical scales are the same in all directions).}
\end{figure}

\end{document}